\documentclass[twocolumn,secnumarabic,amssymb,nobibnotes,aps,prd]{revtex4-1}
\usepackage{graphicx}
\usepackage{color}
\usepackage{alltt}

\makeatletter

\providecommand{\LyX}{L\kern-.1667em\lower.25em\hbox{Y}\kern-.125emX\@}
\DeclareRobustCommand*{\lyxarrow}{%
\@ifstar
{\leavevmode\,$\triangleleft$\,\allowbreak}
{\leavevmode\,$\triangleright$\,\allowbreak}}

\@ifundefined{textcolor}{}
{%
 \definecolor{BLACK}{gray}{0}
 \definecolor{WHITE}{gray}{1}
 \definecolor{RED}{rgb}{1,0,0}
 \definecolor{GREEN}{rgb}{0,1,0}
 \definecolor{BLUE}{rgb}{0,0,1}
 \definecolor{CYAN}{cmyk}{1,0,0,0}
 \definecolor{MAGENTA}{cmyk}{0,1,0,0}
 \definecolor{YELLOW}{cmyk}{0,0,1,0}
}

\makeatother

\begin{document}

\preprint{This line is only printed with preprint option. Vers\~{a}o 1 -- equa\c{c}\~{o}es completas. Modificado em \today}

\title{ Temperature dependence of the magnetic hyperfine field at an s-p impurity diluted in $R$Ni$_{2}$ }

\author{A. L. de Oliveira}
	\email{alexandre.oliveira@ifrj.edu.br}
	\affiliation{Instituto Federal de Educa\c{c}\~{a}o, Ci\^{e}ncia e Tecnologia do Rio de Janeiro, Campus Nil\'{o}polis, Rua L\'{u}cio Tavares 1045, 26530-060, Nil\'{o}polis, RJ, Brazil}

\author{C. M. Chaves}
	\email{cmch@cbpf.br}
	\affiliation{Centro Brasileiro de Pesquisas F\'{\i}sicas, Rua Dr. Xavier Sigaud 150, 22290-180, Rio de Janeiro, RJ, Brazil}

\author{N. A. de Oliveira}
	\affiliation{Instituto de F\'{\i}sica Armando Dias Tavares, Universidade do Estado do Rio de Janeiro, Rua S\~{a}o Francisco Xavier 524, 20550-013, Rio de Janeiro, Brazil}

\author{A. Troper}
	\affiliation{Centro Brasileiro de Pesquisas F\'{\i}sicas, Rua Dr. Xavier Sigaud 150, 22290-180, Rio de Janeiro, RJ, Brazil}

\begin{abstract} 
We study the formation of local magnetic moments and magnetic hyperfine fields at an \textit{s-p} impurity diluted in intermetallic Laves phases compounds  $R$Ni$_{2}$ ($R$ = Nd, Sm, Gd, Tb, Dy) at finite temperatures. We start with a clean host and later the impurity is introduced. The host  has  two-coupled ($R$ and Ni) sublattice Hubbard Hamiltonians but  the Ni sublattice can be disregarded because its $\textit{d}$ band, being full, is magnetically ineffective. Also, the effect of the $\textit{4f}$ electrons of $R$ is represented  by a  polarization of the  \textit{d} band that would be produced by the $\textit{4f}$ magnetic field.This leaves us with a lattice of \textit{effective} rare earth  \textit{R}-ions with only \textit{d} electrons. 
For the \textit{dd} electronic interaction we use the  Hubbard-Stratonovich identity in a functional integral approach in the static saddle point approximation. 
\end{abstract}

\maketitle

\section{introduction}

The Laves phases intermetallic compounds  $R$Ni$_2$ ($R$ = rare earth elements ) crystallize  in a cubic structure~\cite{Buschow77} and exhibit an interesting variety of behaviors related to the changes in their magnetic, electronic, and lattice structures. They  exhibit magnetization associated both with the localized spins ($4f$) and with the itinerant electrons of the rare earth. Although the extensive studies made, the theoretical description of how some of their properties varies with temperature,  due to the electronic correlations, still remains open. The Ni sublattice however can be disregarded because its $\textit{d}$ band, being full, is magnetically ineffective. Also the effect of the $R$ \textit{4f} electrons  is represented by a  polarized  \textit{d} band  produced by the $\textit{4f}$ electrons  magnetic field. This approximation generates an effective rare earth \textit{R} lattice  but whose  interactions differ from the usual $R$ pure metal because now the equivalent lattice constant is different (see the end of section $\bf{III}$ for a numerical indication of the  difference between the two cases).

\section{Method}


 
\subsection{The effective host}

We start with a clean effective host described by the Hamiltonian of itinerant \textit{d}-electrons:
\begin{equation}
\mathcal{H}=H_{d}^{0}+H_{d}^{1}.  \label{eq6:hamini}  
\end{equation}

The first term of Eq.(\ref{eq6:hamini}) is
\begin{equation}
H_{d}^{0} =\sum_{l\sigma }\varepsilon _{0\sigma }d_{l\sigma }^{\dagger}d_{l\sigma }
+\sum_{ll^{\prime }\sigma }T_{ll^{\prime }}d_{l\sigma }^{\dagger }d_{l^{\prime }\sigma },  \label{eq6:h0d}
\end{equation}
where $ \varepsilon _{0\sigma }$ is the energy of the center of the \textit{d} band, now depending on the spin polarization; $d_{j\sigma }^{\dagger }$ ($d_{j\sigma }$) is creation (annihilation) operators, 
and $T_{ll^{\prime}}$ is the hopping integrals between atoms from the $R$ effective lattice. 
  
$H_{d}^{1}$ represents the Coulomb interaction
\begin{equation}
H_{d}^{1}=U\sum_{l}n_{l\uparrow }n_{l\downarrow },
\end{equation}
 $n_{l\sigma }$ being the number operator.


The partition function $\mathcal{Z}$  for the system described by  Hamiltonian (\ref{eq6:hamini})  can be written as
\begin{eqnarray}
\mathcal{Z}& =&\int \prod_{l}\,\mathrm{d}\nu _{l}\,\mathrm{d}\xi _{l}\int
\prod_{j}\,\mathrm{d}\nu _{j}\,\mathrm{d}\xi _{j}\,e^{^{-\beta \left[ -\frac{U}{4}\sum_{l}\left( \nu _{l}^{2}+\xi _{l}^{2}\right)  \right] }}  \nonumber
\\
&& \times \mathrm{tr}\,e^{-\beta \left( 
H_{d}^{0}-\frac{U}{2}
\sum_{l\sigma }\left( i\nu _{l}+\sigma \xi _{l}\right) n_{l\sigma }\right)}\cdot 
\label{Z}
\end{eqnarray}
where $\beta =1/k_{\mathrm{B}}T$, $k_{\mathrm{B}}$ is the Boltzman constant and  $T$ the temperature. The Hubbard-Stratonovich identity\cite{HS} was used in order to  linearize the Coulomb interaction  generating two floating fields,  an electric, $\nu _{l }$, and a magnetic, $\xi _{l}$. The static approximation has also  been  performed.

Now, because of the floating fields, the system, although pure, becomes disordered\cite{IF1, IF2}. In Eq.(\ref {Z}) we see that the floating fields  create  site dependent $ \varepsilon _{l_{0}\sigma } $'s  :
\begin{equation}
   \varepsilon _{l_0\sigma }  = \varepsilon _{0\sigma } -\frac{ U}{2} ( i\nu _{l}+\sigma \xi _{l})             
\end{equation}

   We  then adopt the Coherent Potential Approximation\cite{CPA1, CPA2} (CPA) point of view in which  the system is replaced by an ordered one with an uniform self energy  $\Sigma _{\sigma }$ in all sites $l \neq l_{0}$; in $l_{0}$ the energy $\varepsilon_{l_{0}\sigma }$ remains  function of the  fluctuation fields with respect to $\Sigma _{\sigma }$. 
The partition function in Eq (\ref{Z}) is also the partition function for the Hamiltonian below (which will be the one used to implement the CPA)
\begin{equation}
\tilde{\mathcal{H}}=\tilde{H}^{0}+\tilde{H}^{1},  
\label{eq6:hamiti}
\end{equation}
with
\begin{equation}
\tilde{H}^{0} =
\sum_{\sigma }
\left[ \varepsilon _{l_{0}\sigma }-\Sigma _{\sigma } \right]d_{l_{0}\sigma }^{\dagger }d_{l_{0}\sigma } 
= \sum_{\sigma }V_{\sigma }d_{l_{0}\sigma }^{\dagger }d_{l_{0}\sigma },
\label{H0}
\end{equation}
and
\begin{equation}
\tilde{H}^{1} =  \sum_{l\sigma }\Sigma_{\sigma }d_{l\sigma }^{\dagger }d_{l\sigma }      
+\sum_{ll^{\prime }\sigma }T_{ll^{\prime }}d_{l\sigma }^{\dagger }d_{l^{\prime }\sigma}  
\label{H1cpa}
\end{equation}
 
In (\ref{H1cpa})  both $l$ and $l^\prime$  are  $ \neq l_{0}$. Summing  over all possible $l_0$ we arrive at the self-consistency condition to determine $\Sigma _{\sigma }$:
\begin{equation}
\int \mathrm{d}\xi _{l_{0}} \mathrm{d}\nu _{l_{0}}\frac
{V_{\sigma }(\xi _{l_{0}}\nu _{l_{0}})}
{1-V_{\sigma }(\xi _{l_{0}}\nu_{l_{0}})\ g_{l_{0}l_{0}\sigma }(z)}
P(\xi _{l_{0}}\nu _{l_{0}}) = 0.  
\label{eq6:cpa1a}
\end{equation}
where  the probability distribution $P$, is 
\begin{equation}
P(\xi _{l_{0}}\nu _{l_{0}}) = \frac
{e^{-\beta \Psi (\xi _{l_{0}}\nu_{l_{0}})}}
{\int \mathrm{d}\xi _{l_{0}}\,\mathrm{d}\nu _{l_{0}}\,e^{-\beta \Psi (\xi _{l_{0}}\nu _{l_{0}})}},
\end{equation}
and $\Psi$ is  the free energy associated to $\tilde{H}^{0}$,
\begin{equation}
\Psi (\xi _{l_{0}}\nu _{l_{0}})=\frac{U}{4}\left( \xi_{l_{0}}^{2}+\nu _{l_{0}}^{2}\right) +\frac{1}{\pi }\int \mathrm{d}
\varepsilon \ f(\varepsilon )\ \mathrm{Im}\sum_{\sigma }\ln \left[1-V_{\sigma }(\xi _{l_{0}}\nu _{l_{0}})\ g_{l_{0}l_{0}\sigma }(z)
\right],  \label{eq6:elx}
\end{equation}
$f(\varepsilon )$ is the Fermi function, $z=\varepsilon + i\delta $, $\delta \longrightarrow 0^{+}$, and $g_{l_{0}l_{0}\sigma}(z) $ is the Green function for the hamiltonian $\tilde{H}^{0}$.

Then,  the partition function  (\ref{Z}), in the CPA approach is reduced to
\begin{equation}
\mathcal{Z}=\int \prod_{l_{0}}\mathrm{d}\nu _{l_{0}}\mathrm{d}\xi
_{l_{0}}\ e^{-\beta \Psi \left( \nu _{l_{0}}, \xi _{l_{0}}\right) }.
\end{equation}

\subsection{The introduction of a \textit{s-p} impurity}

We now describe the effects caused by the introduction of a \textit{s-p} impurity (say  Cd). We add a potential $ \mathcal{V}_ {0 \sigma}^ {\mathrm{imp}}$ to Equation~\ref{eq6:hamiti}, 
\begin{equation}
\mathcal{V}_{0\sigma }^{\mathrm{imp}}=\left[ \varepsilon _{\sigma }^{\mathrm{imp}}-\Sigma _{\sigma }(z)\right] d_{l_{0}\sigma }^{\dagger }d_{l_{0}\sigma },
\end{equation}
where the impurity energy $\varepsilon^{\mathrm{imp}}_{\sigma}$ is self consistently determined using the Friedel~\cite{Oliveira2003} condition for the charges screening
\begin{equation}
\Delta Z =\Delta Z_{\uparrow}+\Delta Z_{\downarrow},
\end{equation} 
where  $\Delta Z_{\sigma}$ is the total charge difference between  the $\sigma$ conduction electrons of the impurity and the host:
\begin{equation}
\Delta Z_{\sigma }=\ln \left\{ 1-g_{l_{0}l_{0}\sigma }(\epsilon _{\mathrm{F}})\ \left[ \varepsilon _{\sigma }^{\mathrm{imp}}-\Sigma _{\sigma
}(\epsilon _{\mathrm{F}})\right] \right\} .
\end{equation}

Using the Dyson equation, the perturbed Green functions for this problem can be written as
\begin{equation}
G_{l_{0}l_{0}\sigma }(z) = \frac{g_{l_{0}l_{0}\sigma }(z)}{1- g_{l_{0}l_{0}\sigma }(z) \left[ \varepsilon _{\sigma }^{\mathrm{imp}}-\Sigma _{\sigma }(z)\right] }.
\end{equation}
The local density of states for the $\sigma$ spin direction is 
\begin{equation}
\rho _{0\sigma }(\varepsilon )=-\frac{1}{\pi }\mathrm{Im\ }G_{l_{0}l_{0}\sigma
}(z)
\end{equation}
and the local occupation number is
\begin{equation}
n_{0\sigma }=\int_{-\infty }^{\epsilon _{\mathrm{F}}}\rho _{0\sigma
}(\varepsilon )\,f(\varepsilon )\ \mathrm{d}\varepsilon.
\end{equation}
So, the magnetic moment at the impurity site is 
\begin{equation}
\widetilde{m}(0)=\sum_{\sigma }\sigma n_{0\sigma }
\end{equation}
and finally, we calculate the magnetic hyperfine field at the impurity site, assuming that it is proportional to $\widetilde{m}(0)$, via the temperature independent $A(Z_{\mathrm{imp}})$ Fermi-Segr\`{e} contact coupling parameter~\cite{Campbell69}.:
\begin{equation}
B_{hf}=A(Z_{\mathrm{imp}})\widetilde{m}(0),
\end{equation}
\begin{figure}
   	\includegraphics[width=0.80\columnwidth]{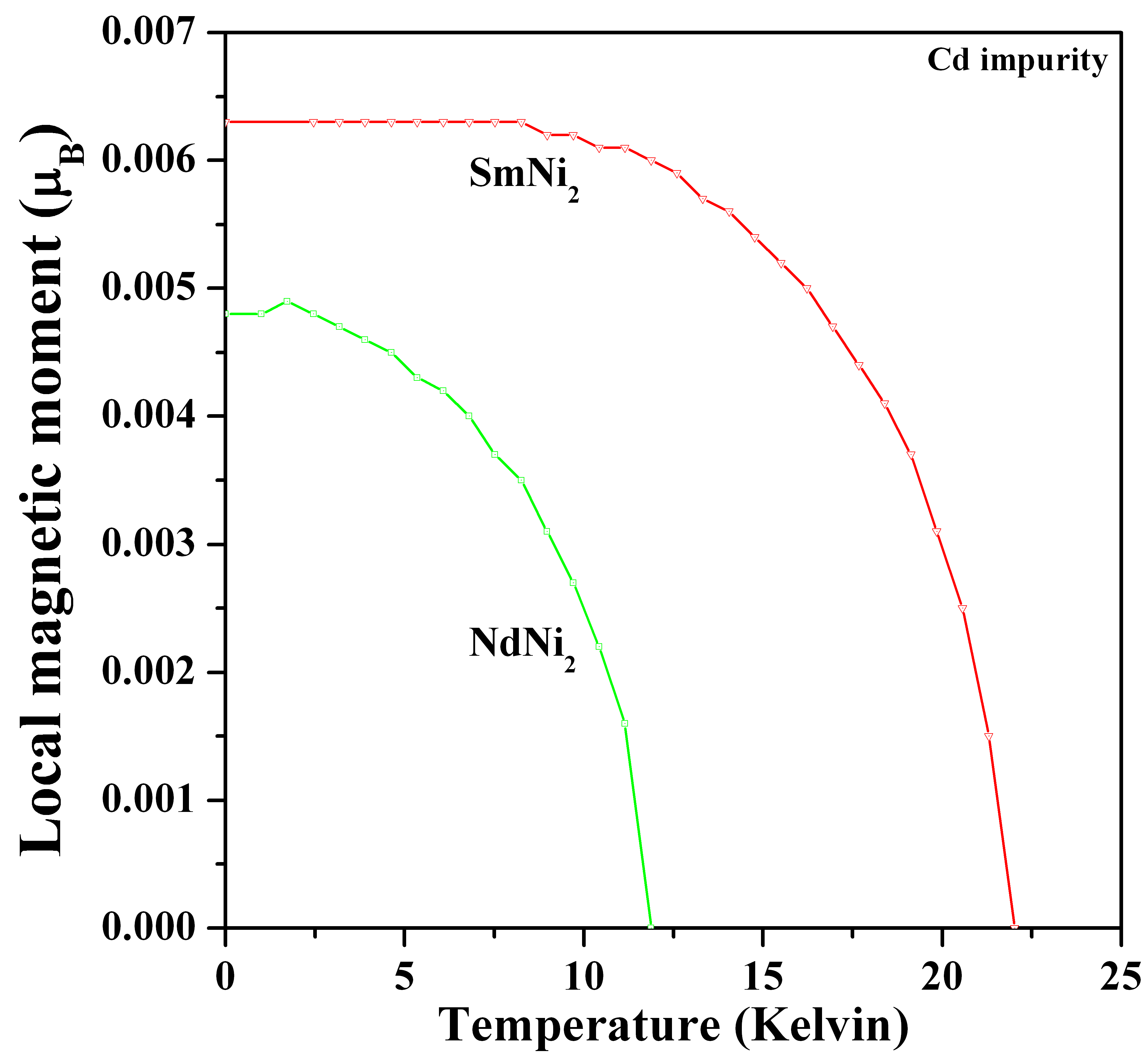}
	\caption{Local magnetic moment at Cd impurity diluted in  $R$Ni$_{2}$ intermetallic host for light  rare earths.}
	\label{fig:LmCdRNi2}
\end{figure}
\begin{figure}
   	\includegraphics[width=0.80\columnwidth]{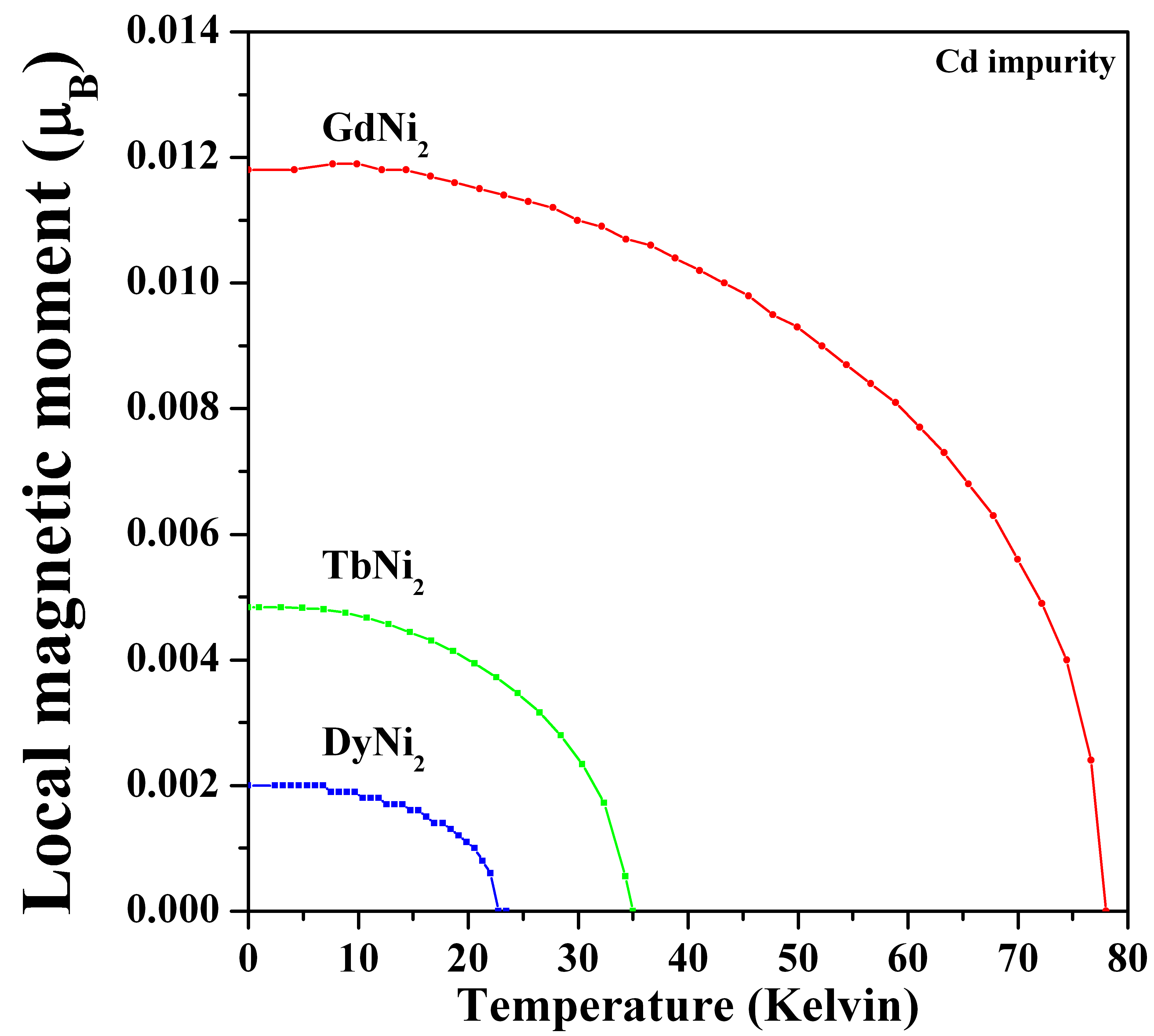}
	\caption{Local magnetic moment at Cd impurity diluted in  $R$Ni$_{2}$ intermetallic host for heavy rare earths.}
	\label{fig:HmCdRNi2}
\end{figure}

\section{Results and Discussions }

We have introduced an effective model to extend our previous zero temperature results~\cite{Oliveira2003} to investigate the magnetic hyperfine fields at a \textit{s-p} impurity such as Cd, in $R$Ni$_{2}$ at finite temperatures. We adopt a standard paramagnetic  density of state extracted from first principle calculation ~\cite{Yama}. For each $R$Ni$_{2}$ compound our model has two adjustable parameters, namely  $\varepsilon_{0\sigma }$ and $U$. These are determined by reproducing the zero  temperature local magnetic moment and the critical temperature.

\begin{figure}
   	\includegraphics[width=0.80\columnwidth]{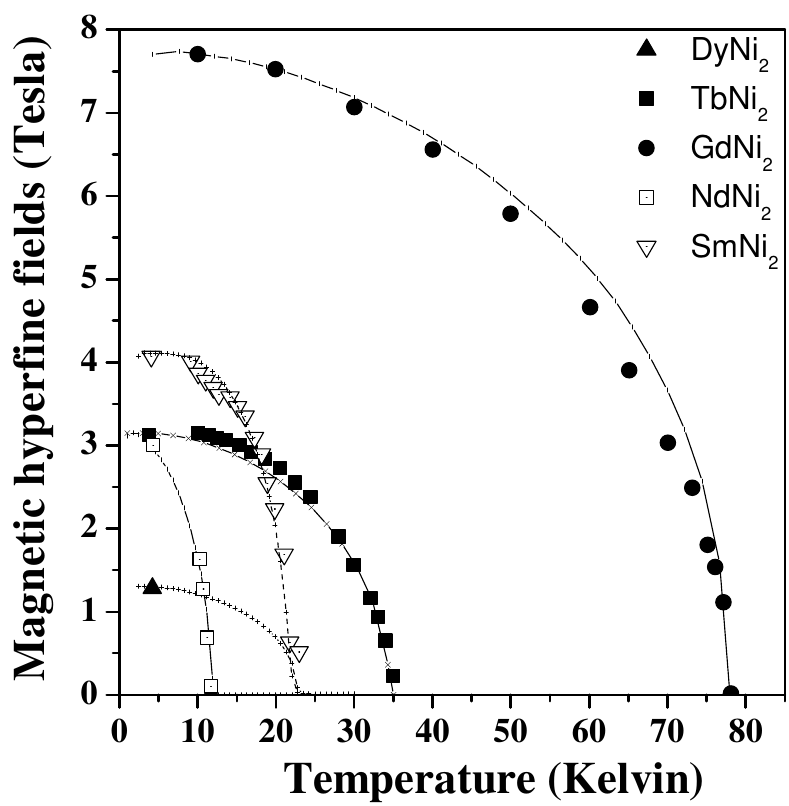}
	\caption{Magnetic hyperfine field at Cd impurity diluted in $R$Ni$_{2}$ intermetallic host. The experimental data were collected from Ref.~\onlinecite{Muller2004}.}
	\label{fig:cdRNi2}
\end{figure}

In Fig.~\ref{fig:LmCdRNi2} , Fig.~\ref{fig:HmCdRNi2} and  Fig.~\ref{fig:cdRNi2} we plot the calculated temperature dependence of the local magnetic moments for the light rare earth elements, for the heavy elements and the magnetic hyperfine fields at Cd  as function of temperature.The results are  good agreement with the  experimental results~\cite{Muller2004}.

As stated before, in the effective $R$ lattice the interactions are different from a $R$ pure metal. From  Fig.~\ref{fig:HmCdRNi2} we see that the local moment, in units of Bohr magneton $\mu_B$, at $T=0K$ is about 0.012 whereas in Ref. \onlinecite{alekos} it was found that in pure Gd metal it is about 0.05.

\begin{acknowledgments}
We would like to aknowledge the support from the Brazilian agencies FAPERJ and CNPq.
\end{acknowledgments}

\end{document}